\documentclass[11pt]{article}

\begin{document}

\baselineskip=16pt


\newcommand{\be}{\begin{equation}}
\newcommand{\ee}{\end{equation}}
\newcommand{\ba}{\begin{eqnarray}}
\newcommand{\ea}{\end{eqnarray}}
\newcommand{\no}{\nonumber \\}

\makeatletter
\renewcommand{\theequation}{\thesection.\arabic{equation}}
\@addtoreset{equation}{section}
\makeatother
\begin{titlepage}
\title{
\hfill\parbox{4cm}
{\normalsize {\tt hep-th/0607123}}\\
\vspace{1cm}
{\bf A Holographic Dual of Hydrodynamics}
}
\author{
Shin Nakamura$\,^{a,b}$
\thanks{{\tt E-mail: nakamura@ihanyang.ac.kr}}
$\ $and$\ $ Sang-Jin Sin$\,^{a}$
\thanks{{\tt E-mail: sjsin@hanyang.ac.kr}}
\vspace{5mm}\\[15pt]
$^a$~{\it Physics Department, Hanyang University, Seoul, 133-791, Korea}\\
\\
$^b$~{\it Center for Quantum Spacetime (CQUeST)}\\
{\it Sogang University, Seoul, 121-742, Korea}\\
\\[10pt]
}
\date{July 19, 2006}
\maketitle
\thispagestyle{empty}

\begin{abstract}
\normalsize
\noindent
We consider a gravity dual description of time dependent,
strongly interacting large-$N_{c}$ ${\cal N}=4$ SYM.
We regard the gauge theory system as a fluid with shear viscosity.
Our fluid is expanding in one direction following the
Bjorken's picture that is relevant to RHIC experiments.
We obtain the dual geometry at the late time that is consistent
with dissipative hydrodynamics.
We show that the integration constants that cannot be determined
by hydrodynamics are given by looking at the horizon of the
dual geometry.
Relationship between time dependence of the energy density
and bulk singularity is also discussed.
\end{abstract}

\end{titlepage}

\section{Introduction}
One of the attractive aspects of AdS/CFT \cite{AdS/CFT} is
applicability to the real systems after certain amount of deformations.
In fact it has been suggested that the fireball in
Relativistic Heavy Ion Collision (RHIC)
can be explained from dual gravity point of view
\cite{SZ,Nastase,SSZ,Aharony}, since the quark-gluon systems
created there are in the strong coupling region \cite{RHIC}.
Although the YM theory described by the standard AdS/CFT
is large-$N_{c}$ ${\cal N}=4$ SYM theory,
there are many attempts to construct models closer to QCD
\cite{AdS/QCD}.
SUSY is not very relevant in the finite temperature context
since it is broken completely.

Since the RHIC fireball is expanding, we need to understand
AdS/CFT in the time dependent situation.
Recently, Janik and Peschanski \cite{Janik, Janik-2} discussed
this problem in non-viscous case.
They use the conservation law and conformal invariance together
with the holographic renormalization \cite{HSS,Skend} to express
the bulk geometry with given boundary data.
As a result, the bulk geometry reproduces the basic features of
Bjorken theory \cite{Bjorken}.
It is also pointed out that inclusion of shear viscosity,
although the value is small, is very important in
the analyses of real RHIC physics since it plays an essential
role in the elliptic flow (see for example, \cite{KH-Chau,bulk-zero}).
In fact, the shear viscosity at the strong coupling limit
was calculated for the ${\cal N}=4$ SYM systems in Ref.
\cite{PSS-1} using AdS/CFT.
So it is natural to ask how the bulk geometry change if we include
the viscous effects in the boundary theory.

In this paper, we establish the dual geometry in the presence of
shear viscosity by using the hydro-dynamics as the boundary data.
Although our gauge theory is not QCD, we hope there is   universal
features in the character of strongly interacting gauge theory
systems. In fact hydrodynamics, which is our input does not ask much
about the details of the microscopic particles and the interactions
once the equation of state is given. Therefore we have a chance to
extract useful information on the macroscopic properties of the real
quark-gluon fluid based on this universality.

If what we get is consistency with fluid dynamics, there would not
be much point to consider AdS/CFT dual of it. In fact, the
holographic dual of the hydrodynamics contains much more information
than the hydrodynamics since AdS/CFT already contains essential
information of microscopic gauge theory dynamics. For example, we
will show that the holographic dual of the hydrodynamics gives
integration constants in the hydrodynamic equations that cannot be
determined by hydrodynamics alone. The dual geometry also gives a
simple derivation of Stefan-Boltzmann's law in strongly coupled
regime with precise Stefan-Boltzmann constant.

The organization of the paper is the following. In Section
\ref{hydro}, we analyze time dependence of the system in the
framework of the relativistic hydrodynamics. We also review the
basics of the dissipative relativistic hydrodynamics in order to
clarify our setup. Section \ref{gravity} gives the analysis in the
gravity dual. We review the basic framework of the gravity dual and
present some results for non-viscous cases obtained in Ref.
\cite{Janik}. The main results of the present work will be given in
Section \ref{viscous} where the late time dual geometry
is proposed and consistency with the hydrodynamic analyses
is checked.
We will show that the holographic dual of the hydrodynamics
contains more information than the hydrodynamics.
We also make comments on the regularity of the bulk
geometry in Section \ref{singularity}.
We conclude in the final section.

\section{Relativistic hydrodynamics with shear viscosity}
\label{hydro}

We begin with a short review of the relativistic hydrodynamics
with dissipation.
For the relevance to the RHIC fireball, we assume that it is
described by the finite temperature theory of
a variant of ${\cal N}=4$ SYM.
We also
follow the Bjorken's picture \cite{Bjorken}. In the Bjorken's model,
the system undergoes  one-dimensional expansion (Bjorken expansion)
along the collision axis of the heavy ions, and the fluid of the
quarks and gluons has boost symmetry in the so-called central
rapidity region \cite{Bjorken}. We shall consider only the late-time
regime of the Bjorken expansion where the time evolution is slow
enough to employ approximations.\footnote{Realistic model should
contain three-dimensional expansion as tried in Ref. \cite{SSZ}. 
There, it was suggested to use a dual of three-dimensional 
cosmic expansion.}

The energy-momentum tensor in the framework of relativistic
hydrodynamics is known to be\footnote{
The convention of the signature of the metric is
$(-,+,+,+)$ in this paper.}
\begin{eqnarray}
T^{\mu\nu}=(\rho+P)u^{\mu}u^{\nu}+Pg^{\mu\nu}+\tau^{\mu\nu},
\label{T-ideal}
\end{eqnarray}
where $\rho$, $P$ are the energy density and the pressure of
the fluid, and $u^{\mu}=(\gamma, \gamma \vec{v})$ is the four-velocity
field in terms of the local fluid velocity $\vec{v}$.
$\tau^{\mu\nu}$ is the dissipative term.
In a frame where the energy three-flux vanishes,
$\tau^{\mu\nu}$ is given in terms of the bulk viscosity
$\xi$ and the shear viscosity $\eta$ by
\begin{eqnarray}
\tau^{\mu\nu}
=-\eta
(\bigtriangleup^{\mu\lambda}\nabla_{\lambda}u^{\nu}
+\bigtriangleup^{\nu\lambda}\nabla_{\lambda}u^{\mu}
-\frac{2}{3}\bigtriangleup^{\mu\nu}\nabla_{\lambda}u^{\lambda})
-\xi \bigtriangleup^{\mu\nu}\nabla_{\lambda}u^{\lambda},
\label{T-dissp}
\end{eqnarray}
under the assumption that $\tau^{\mu\nu}$
is of first order in gradients.
We have defined the three-frame projector as
$\bigtriangleup^{\mu\nu}=g^{\mu\nu}+u^{\mu}u^{\nu}$.

In this paper, we consider pure ${\cal N}=4$ SYM theory
whose energy-momentum tensor is traceless.
Now, the trace of the energy-momentum tensor is given by
\begin{eqnarray}
T^{\mu}_{\mu}=-\rho+3P-3\xi\nabla_{\lambda}u^{\lambda}.
\end{eqnarray}
Demanding $T^{\mu}_{\mu}=0$ for all the possible
frames where (\ref{T-dissp}) is valid,
we obtain
\begin{eqnarray}
 \xi=0, ~~~{\rm and}~~~ \rho=3P.
\label{eos}
\end{eqnarray}
Notice that the bulk viscosity in the realistic RHIC setup might
also be negligible. (See for example, Ref. \cite{bulk-zero}.)

We assume that our fluid system
is boost invariant following Bjorken \cite{Bjorken}
since it is actually supported by experiments.
We want to take a ``co-moving frame'' where
each point of the fluid labels the coordinate, a concept called
Lagrangian frame in fluid dynamics.
In this frame all the fluid points are at rest by definition,
hence  all the fluid points share the same proper time.
We can use  the rapidity of each fluid-point as a spatial
coordinate and the common proper time of each fluid-point
as a time coordinate.
Therefore a local rest frame (LRF) of the fluid
can be given by proper time($\tau$)-rapidity($y$), whose
relationship with the
cartesian coordinate is
$(x^{0},x^{1},x^{2},x^{3})=(\tau \cosh y,\tau \sinh y,x^{2},x^{3})$.
We have chosen the collision axis to be in the $x^{1}$ direction.

The Minkowski metric in this coordinate has the form of
\begin{eqnarray}
ds^{2}=-d\tau^{2}+\tau^{2}dy^{2}+dx_{\perp}^{2},
\label{g00}
\end{eqnarray}
where $dx_{\perp}^{2}=(dx^{2})^{2}+(dx^{3})^{2}$.
We assume that the collision happened at $\tau=0$ and we consider
only $\tau\ge 0$ region. Note that $|y|\sim  \infty$ corresponds to
the fronts of the expanding fluid. Therefore, the whole region on
the y-coordinate axis is occupied by the fluid. We also assume that
the fluid  is extended in the $x^{2}, x^{3}$ directions
homogeneously. Since  the real fireball produced by RHIC experiment
is localized, the set up we use is an idealized one. Nevertheless,
the present setup is proper since we are interested in the central
rapidity region.

The four-velocity of the fluid at any point in  the LRF
is $u^{\mu}=(1,0,0,0)$, and
this makes the energy-momentum tensor to be diagonal:
\begin{eqnarray}
T^{\mu\nu}
=
\left(
  \begin{array}{cccc}
   \rho&   0&   0&   0\\
      0&\frac{1}{\tau^{2}}
      \left(P-\frac{4}{3}\frac{\eta}{\tau}\right)
      &   0&   0\\
      0&   0&P+\frac{2}{3}\frac{\eta}{\tau}&   0\\
      0&   0&   0&P+\frac{2}{3}\frac{\eta}{\tau}\\
  \end{array}
\right).
\label{T-diag}
\end{eqnarray}
We have three independent quantities, $\rho$, $P$ and $\eta$
in (\ref{T-diag}). However,
the energy-momentum conservation, $\nabla_{\mu}T^{\mu\nu}=0$,
together with the equation of state $\rho=3P$, reduces the number
of the independent quantities to be one.
One finds that the energy-momentum tensor is written by using
only $\rho$ in the following way:
\begin{eqnarray}
T^{\mu\nu}
=
\left(
  \begin{array}{cccc}
   \rho&   0&   0&   0\\
      0&\frac{1}{\tau^{2}}
      \left(-\rho-\tau \dot{\rho} \right)
      &   0&   0\\
      0&   0&\rho+\frac{1}{2}\tau\dot{\rho}&   0\\
      0&   0&   0&\rho+\frac{1}{2}\tau\dot{\rho}\\
  \end{array}
\right),
\label{T-diag-2}
\end{eqnarray}
where $\dot{\rho}\equiv \frac{d \rho}{d\tau}$.
By identifying (\ref{T-diag}) with (\ref{T-diag-2}), we
 obtain the following differential equation that connects
$\eta$ and $\rho$:\footnote{
The the equation (\ref{diff-eq}) turns out to be the same
as the one appearing in so-called first order (or standard)
dissipative relativistic hydrodynamics.
(See for example, Ref. \cite{Muronga} and the cited therein.)
It is known that the first order formalism has a problem
of acausal signal propagation.
However, it gives good enough results for our purposes.
The details of the causal dissipative relativistic
hydrodynamics and consistency of our analysis are shown
in Appendix \ref{second-from}. }

\begin{eqnarray}
\frac{d\rho}{d\tau}
=-\frac{4}{3}\frac{\rho}{\tau}
+\frac{4}{3}\frac{\eta}{\tau^{2}}.
\label{diff-eq}
\end{eqnarray}
Note that both of $\rho$ and $\eta$ depend on the proper time
$\tau$ in general.
Let's assume that the shear viscosity evolves by
\begin{eqnarray}
\eta=\frac{\eta_{0}}{\tau^{\beta}},
\label{eta}
\end{eqnarray}
where $\eta_{0}$ is a positive constant.
The solution of (\ref{diff-eq}) is then given by
\begin{eqnarray}
\rho(\tau)
&=&\frac{\rho_{0}}{\tau^{4/3}}
+\frac{4\eta_{0}}{1-3\beta} \frac{1}{\tau^{1+\beta}}
\:\:\:\:({\rm for}\:\: \beta\neq1/3),
\label{solution}\\
\rho(\tau) &=&\frac{\rho_{0}}{\tau^{4/3}}
+\frac{4\eta_{0}}{3} \frac{\ln(\tau)}{\tau^{4/3}}
\:\:\:\:({\rm for}\:\: \beta=1/3),
\label{critical-solution}
\end{eqnarray}
where $\rho_{0}$ is a positive constant.
For $\beta \leq 1/3$ case,
the viscous corrections in the hydrodynamic quantities become
dominant in the late time, which invalidates the hydrodynamic
description. If $\beta>\frac{1}{3}$, the shear viscosity term
is sub-leading in the late time behavior as we expect. Therefore we
will consider only $\beta>\frac{1}{3}$ case from now on.

The proper time dependence of the temperature $T$
can be read off by assuming
the Stefan-Boltzmann's law $\rho \propto T^{4}$:
\begin{eqnarray}
T=
T_{0}\left(
\frac{1}{\tau^{1/3}}
+
\frac{\eta_{0}}{\rho_{0}}\frac{1}{1-3\beta}
\frac{1}{\tau^{\beta}}
+\cdots
\right).
\label{T-general}
\end{eqnarray}
In the {\em static} finite temperature system of strongly coupled
${\cal N}=4$ SYM theory, it is known that
$\eta \propto T^{3}$ \cite{PSS-1}.
Let us assume that the same is true in the slowly varying
non-static cases. Then we set $\beta=1$:
\begin{eqnarray}
\eta=\frac{\eta_{0}}{\tau}.
\label{beta1}
\end{eqnarray}
We know $\rho\sim T^4$ and $\eta\sim T^{3}$ cannot be
consistent without an additional term in (\ref{beta1}),
but the correction term is negligible in our case.
One can check the consistency of our approach
in Appendix \ref{consistency}.
The temperature behavior is then given by
\begin{eqnarray}
T=T_{0}\left(
\frac{1}{\tau^{1/3}}
-
\frac{\eta_{0}}{2\rho_{0}}
\frac{1}{\tau}
+\cdots
\right).
\end{eqnarray}

We can evaluate the entropy change in the presence of shear
viscosity by using hydrodynamics.
The conservation of energy-momentum tensor can be rewritten as
\begin{eqnarray}
\frac{d(\tau\rho)}{d\tau}+P = \frac{4\eta}{3\tau}.
\end{eqnarray}
Using the nature of the one-dimensional expansion, the above
can be rephrased as
\begin{eqnarray}
T\frac{d(\tau s)}{d\tau}=\frac{4\eta}{3\tau},
\end{eqnarray}
where $s$ denotes the entropy density and $\tau s \equiv S$
is the entropy per unit rapidity and unit transverse
area.\footnote{
A precise definition of $S$ is the entropy within
a unit 3d region on the $(y, x^{2}, x^{3})$ coordinate.
The volume of this region is $\tau$ and it is expanding with
time in the $x^{1}$ direction.
}
Notice that in the absence of viscosity, $S$ is constant.
Now, the entropy per unit rapidity and unit transverse area
has time dependence,
\begin{eqnarray}
S(\tau)
&=&\int d\tau \frac{4\eta}{3\tau T}
\nonumber \\
&=&
S_{\infty}-2\frac{\eta_{0}}{T_{0}}\tau^{-2/3}
+ O(\tau^{-4/3}),
\label{entropy}
\end{eqnarray}
due to creation of the entropy by dissipation.
However, the creation rate of the entropy slows down with time.
$S_{\infty}$, that is the entropy per unit rapidity and unit
transverse area at $\tau=\infty$, is an integration constant
which we cannot determine in the framework of hydrodynamics.
We will show that its precise value is given by using the
gravity dual in Section \ref{gravity}.

\section{Holographic Dual of Hydrodynamics}
\label{gravity}

In this section, we will find a five-dimensional metric
which is dual to the hydrodynamic description of the YM fluid
in the previous section.
The basic strategy is to use the Eistein's equation
together with the boundary condition given by the
energy-momentum tensor at the boundary \cite{HSS,Skend,Janik}.
We consider general asymptotically AdS metrics in the
Fefferman-Graham coordinate:
\begin{eqnarray}
ds^{2}=
r_{0}^{2}
\frac{g_{\mu\nu}dx^{\mu}dx^{\nu}+dz^{2}}{z^{2}},
\label{FG-metric}
\end{eqnarray}
where $x^{\mu}=(\tau,y,x^{2},x^{3})$ in our case.
$r_{0}\equiv (4\pi g_{s} N_{c} \alpha'^{2})^{1/4}$
is the length scale given by the string coupling $g_{s}$ and
the number of the colors $N_{c}$.
The four-dimensional metric $g_{\mu\nu}$ is expanded
with respect to $z$ in the following form \cite{HSS,Skend}:
\begin{eqnarray}
g_{\mu\nu}(\tau, z)=
g^{(0)}_{\mu\nu}(\tau)+z^{2}g^{(2)}_{\mu\nu}(\tau)
+z^{4}g^{(4)}_{\mu\nu}(\tau)
+z^{6}g^{(6)}_{\mu\nu}(\tau)+\cdots .
\label{expansion}
\end{eqnarray}
$g^{(0)}_{\mu\nu}$ is the physical four-dimensional metric for the
gauge theory on the boundary, that is given by (\ref{g00}) in
the present case.
The $g_{\mu\nu}^{(n)}$'s depend only on $\tau$
because of the translational symmetry in the $x^2$, $x^3$
directions and the boost symmetry in the $y$ direction in our setup.
$g^{(2)}_{\mu\nu}$ is found to be zero. We can identify
the first non-trivial data in (\ref{expansion}), $g^{(4)}_{\mu\nu}$,
with the energy-momentum tensor at the boundary \cite{HSS}:
\begin{eqnarray}
g^{(4)}_{\mu\nu}
=\frac{4\pi G_{5}}{r_{0}^{3}} \langle T_{\mu\nu} \rangle,
\end{eqnarray}
where $G_{5}$ is the 5d Newton's constant given by
$G_{5}=8\pi^{3}\alpha'^{4}g_{s}^{2}/r_{0}^{5}$ in our notation.
For the time being, we set $4\pi G_{5}=1$ and $r_{0}=1$.
The higher-order terms in (\ref{expansion}) are determined by
solving the Einstein's equation with negative cosmological
constant $\Lambda=-6$
\cite{HSS, Janik}:
\begin{eqnarray}
R_{MN}-\frac{1}{2}G_{MN}R-6G_{MN}=0,
\label{Eeq}
\end{eqnarray}
where the metric and the curvature tensor are for the five-dimensional
ones of (\ref{FG-metric}).
$g_{\mu\nu}^{(2n)}$ is described by
$g_{\mu\nu}^{(2n-2)}, g_{\mu\nu}^{(2n-4)}, \cdots, g_{\mu\nu}^{(0)}$
through solving the Einstein's equation.
In other words, we
can obtain the higher-order terms in (\ref{expansion})
recursively by starting with the initial data
$g_{\mu\nu}^{(0)}$ ($\sim$Minkowski)
and $g_{\mu\nu}^{(4)}$ ($\sim T_{\mu\nu}$).

\subsection{Static cases}

In order to demonstrate the use of the above procedure,
we first workout the static case using cartesian coordinate
$x^{\mu}=(t,x^{1},x^{2},x^{3})$ instead of the Bjorken's coordinate.
The energy-momentum tensor in this case is given by
\begin{eqnarray}
T_{\mu\nu}
={\rm diag}
\left(\rho, \rho/3, \rho/3, \rho/3\right).
\label{Tij-static}
\end{eqnarray}
The result of the above procedure gives the solution of the
Einstein's equation in Fefferman-Graham co-ordinate:
\begin{eqnarray}
ds^{2}=
\frac{1}{z^{2}}
\left\{
-\frac{(1-\frac{\rho}{3}z^{4})^{2}}
{1+\frac{\rho}{3}z^{4}}
dt^{2}
+
\left(1+\frac{\rho}{3}z^{4}\right)
(dx_{1}^{2}+dx_{2}^{2} +dx_{3}^{2})
\right\}
+\frac{dz^{2}}{z^{2}},
\label{AdS-BH}
\end{eqnarray}
which is equivalent to the AdS-Schwarzschild
Black hole.\footnote{Notice that the metric is mapped to the
standard form of the AdS-Schwarzschild metric through the
coordinate transformation
$\tilde{z}=z/\sqrt{1+z^{4}/z^{4}_{0}}$.}
The Hawking temperature is given by
\begin{eqnarray}
T_{H}=\sqrt{2}/(z_{0}\pi),
\end{eqnarray}
where $z_{0}(\tau)=[3/\rho]^{1/4}$ is the position of the horizon.
By restoring $4\pi G_{5}$ and
$r_{0}\equiv (4\pi g_{s} N_{c} \alpha'^{2})^{1/4}$,
and by identifying $T_{H}$ with gauge theory temperature $T$,
we obtain the Stefan-Boltzmann's law\footnote{
Actually Stefan Boltzmann's law is about intensity
$I=c\rho/4$ in terms of temperature. But we use the
terminology abusively to name (\ref{SB}).}
\begin{eqnarray}
\rho=\frac{r_{0}^{3}}{4\pi G_{5}}\frac{3\pi^{4}}{4}T_{H}^{4}
=
\frac{3}{8}\pi^{2}N_{c}^{2}T ^{4}.
\label{SB}
\end{eqnarray}
This result agrees with  Ref. \cite{GKPeet}.

\subsection{Non-viscous time dependent cases}
\label{non-vis}

Coming back to Bjorken's setup, Janik and Peschanski
obtained the late time bulk metric \cite{Janik}
by the above procedure. Here we briefly review their work
in our language.
By using (\ref{T-diag-2}) and (\ref{solution}),
the energy-momentum tensor for non-viscous case is explicitly
written as
\begin{eqnarray}
T_{\mu\nu}=\left(
  \begin{array}{cccc}
    \frac{\rho_{0}}{\tau^{4/3}}
       & 0  & 0  & 0  \\
     0 & \tau^{2}\frac{\rho_{0}}{3\tau^{4/3}}
       & 0  & 0  \\
     0 & 0  & \frac{\rho_{0}}{3\tau^{4/3}} & 0  \\
     0 & 0  & 0  & \frac{\rho_{0}}{3\tau^{4/3}}\\
  \end{array}
\right).
\label{Tij-nonviscos}
\end{eqnarray}
Then the metric is given by
\begin{eqnarray}
g_{\tau\tau}&=&-1+\frac{\rho_{0}}{\tau^{4/3}}z^{4}+O(z^{6}),
\nonumber \\
\frac{g_{yy}}{\tau^{2}}&=&1+\frac{\rho_{0}}{3\tau^{4/3}}z^{4}+O(z^{6}),
\nonumber \\
g_{xx}&=&1+\frac{\rho_{0}}{3\tau^{4/3}}z^{4}+O(z^{6}),
\end{eqnarray}
where $g_{xx}=g_{22}=g_{33}$.
Notice that our Minkowski metric is given by (\ref{g00}).
Let's focus  on the late time behaviour
of the metric, since $T_{\mu\nu}$ is given only for the
late time.\footnote{
The slow time evolution is necessary to justify the
hydrodynamic treatment of the fluid.}
If we take $\tau \to \infty$ limit naively,
what we obtain is just the Minkowski metric
(\ref{g00}).
To extract a non-trivial result, the authors of Ref. \cite{Janik}
take the limit such that $g^{(4)}z^{4}$ does not go to zero nor
infinity as $\tau \to \infty$:
\begin{eqnarray}
\tau \to \infty \:\:\:\:\:\:{\rm with}\:\:
\frac{z}{\tau^{1/3}}\equiv v \:\:{\rm fixed}.
\label{Janik-limit}
\end{eqnarray}
Then by solving the Einstein's equation recursively up to certain
order of $z$,  $g_{\tau\tau}$,
$g_{yy}/\tau^{2}$ and $g_{xx}$ have the following structure:
\begin{eqnarray}
f^{(1)}(v)+f^{(2)}(v)/\tau^{4/3}+\cdots.
\label{expansion1}
\end{eqnarray}
By neglecting the $O(\tau^{-4/3})$ quantities, they
obtained an analytic expression of the late time metric:
\begin{eqnarray}
ds^{2}=
\frac{1}{z^{2}}
\left\{
-\frac{(1-\frac{\rho_{0}}{3}\frac{z^{4}}{\tau^{4/3}})^{2}}
{1+\frac{\rho_{0}}{3}\frac{z^{4}}{\tau^{4/3}}}
d\tau^{2}
+
\left(1+\frac{\rho_{0}}{3}\frac{z^{4}}{\tau^{4/3}}\right)
(\tau^{2}dy^{2}+dx^{2}_{\bot})
\right\}
+\frac{dz^{2}}{z^{2}}.
\label{Janik-bulk}
\end{eqnarray}
We can explicitly check that the above metric satisfies
\begin{eqnarray}
G^{LM}(R_{MN}-\frac{1}{2}G_{MN}R-6G_{MN})
%
\sim O(1/\tau^{2}).
\label{EEQ}
\end{eqnarray}
Notice that (\ref{Janik-bulk}) is a black hole in AdS space with
time-dependent horizon. The time dependence of the entropy and the
Hawking temperature from the metric (\ref{Janik-bulk}) reproduces
the Bjorken's results \cite{Bjorken} $S\sim constant$ as well as
$T\sim \tau^{-1/3}$. Interestingly, they observed that the
regularity of the geometry at the horizon uniquely select the power
of time evolution of energy density $\rho\sim \tau^{-4/3}$, which is
a consequence of hydrodynamics. (See Section \ref{singularity} for
the details.)

If we replace $\rho=\frac{\rho_{0}}{3}\frac{1}{\tau^{4/3}}$
with $\rho=\frac{\rho_{0}}{3}\frac{\log\tau}{\tau^{4/3}}$,
we find that the left-hand-side of (\ref{EEQ}) is at the order
of $1/(\tau^{2}\log\tau)$.
In this sense, this replacement makes another late time
solution of (\ref{Eeq}). We will use this solution in Section
\ref{singularity}.

\subsection{Viscous cases}
\label{viscous}

Let us come back to our main interest  to obtain the bulk geometry in the presence of shear viscosity.
The energy-momentum tensor for $\beta=1$ is
written by using (\ref{T-diag-2}) and (\ref{solution}) as
\begin{eqnarray}
T_{\mu\nu}=\left(
  \begin{array}{cccc}
    \frac{\rho_{0}}{\tau^{4/3}}
    -\frac{2\eta_{0}}{\tau^{2}}
       & 0  & 0  & 0  \\
     0 & \tau^{2}\left(\frac{\rho_{0}}{3\tau^{4/3}}
    -\frac{2\eta_{0}}{\tau^{2}}\right)   & 0  & 0  \\
     0 & 0  & \frac{\rho_{0}}{3\tau^{4/3}} & 0  \\
     0 & 0  & 0  & \frac{\rho_{0}}{3\tau^{4/3}}\\
  \end{array}
\right).
\label{Tij-viscous}
\end{eqnarray}
The metric components,
$g_{\tau\tau}$, $g_{yy}/\tau^{2}$, $g_{xx}$
have the following structure by solving the Einstein's equation
recursively:
\begin{eqnarray}
f^{(1)}(v)+\eta_{0}h^{(1)}(v)/\tau^{2/3}+
{\tilde f}^{(2)}(v)/\tau^{4/3}+\cdots ,
\label{expansion2}
\end{eqnarray}
Note that {\it the viscosity dependent terms exist at the
order of $\tau^{-2/3}$ and these are
more important than the higher-order terms neglected
in (\ref{Janik-bulk})}.
We are considering the late time region $\tau \gg 1$.
But to see the effects of viscosity,
we need to keep the terms at least to the order of $\tau^{-2/3}$.
In this paper we consider the viscosity effects to the minimal order.

Now we solve the Einstein's equation recursively.
The power series that appear in the solution
can be re-summed to give a compact form of the metric.
For the detail, see Appendix \ref{derivation}.
The late time 5d bulk geometry is given by
\begin{eqnarray}
ds^{2}
&=&
\frac{1}{z^{2}}
\left\{
-\frac{(1-\frac{\rho z^{4}}{3})^{2}}{1+\frac{\rho z^{4}}{3}}
d^{2}\tau
+
\left(1+\frac{\rho z^{4}}{3}\right)
\left(
\frac{1+\frac{\rho z^{4}}{3}}{1-\frac{\rho z^{4}}{3}}
\right)^{-2\gamma}
\tau^{2}d^{2}y
+
\left(1+\frac{\rho z^{4}}{3}\right)
\left(
\frac{1+\frac{\rho z^{4}}{3}}{1-\frac{\rho z^{4}}{3}}
\right)^{\gamma}
d^{2}x_{\bot}
\right\}
\nonumber \\
&&+\frac{dz^{2}}{z^{2}},
\label{our-geometry}
\end{eqnarray}
where
\begin{eqnarray}
\gamma \equiv \frac{\eta_{0}}{\rho_{0}\tau^{2/3}}
 ~~{\rm and}~~
\rho=
\frac{\rho_{0}}{\tau^{4/3}}
-
\frac{2\eta_{0}}{\tau^{2}}.
\label{def-gamma-1}
\end{eqnarray}
Notice that the energy momentum tensor (\ref{Tij-viscous}) can NOT
be written in terms of the whole $\rho(\tau)$. It is truly amazing
that the final metric nevertheless can be written in terms of
$\rho(\tau)$  (apart from the power) in the compact way. This
implies that the position of horizon can be determined solely by the
energy density.\footnote{ One should keep in mind that we are
looking for the late time geometry; the metric (\ref{our-geometry})
is correct only to the order of $\gamma$ and the $O(\gamma^2)$
contributions are not unambiguously determined. The representation
of (\ref{our-geometry}) is chosen since it makes the volume of the
horizon finite.}

The Hawking temperature in the adiabatic approximation
is given by $T(\tau)=\sqrt{2}/\pi(z_{0}(\tau))$,
where $z_{0}(\tau)=[3/\rho(\tau)]^{1/4}$ is the time dependent
position of the horizon. Just as the static case, we obtain
\begin{eqnarray}
\rho=
\frac{3}{8}\pi^{2}N_{c}^{2}T^{4}(\tau),
\end{eqnarray}
by restoring $4\pi G_{5}$ and $r_{0}$.
The entropy per unit rapidity and unit transverse area
is given by
\begin{eqnarray}
S
&=&\frac{1}{4G_{5}} \frac{2\sqrt{2}\tau r_{0}^{3}}{z_{0}^{3}(\tau)}
\nonumber \\
&=&
\left(\frac{N_{c}^{2}}{2\pi}\right)^{1/4}
\left(\frac{\pi}{3}\right)^{3/4}
2\sqrt{2}\rho_{0}^{3/4}
\left(
1-\frac{3}{2}\frac{\eta_{0}}{\rho_{0}\tau^{2/3}}
+O(\tau^{-4/3})
\right).
\label{s-evolve-ads}
\end{eqnarray}
One remarkable thing is that
the value of $S$ at $\tau=\infty$, that cannot be determined by
hydrodynamics alone, is precisely determined to be
\begin{eqnarray}
S_{\infty}
=\left(\frac{N_{c}^{2}}{2\pi}\right)^{1/4}
\left(\frac{\pi}{3}\right)^{3/4}
2\sqrt{2}\rho_{0}^{3/4},
\end{eqnarray}
in terms of the initial condition $\rho_{0}$.

Let us check consistency of (\ref{s-evolve-ads}) and (\ref{entropy}).
The normalized entropy-creation rate is given by
\begin{eqnarray}
\frac{1}{S}\frac{dS}{d\tau}
=\frac{\eta_{0}}{\rho_{0}\tau^{5/3}}+O(\tau^{-7/3})
\label{rate-ads}
\end{eqnarray}
from the gravity dual and
\begin{eqnarray}
\frac{1}{S}\frac{dS}{d\tau}
=\frac{4}{3}\frac{\eta_{0}}{T_{0}S_{\infty}\tau^{5/3}}
+O(\tau^{-7/3})
\label{rate-hydro}
\end{eqnarray}
from (\ref{entropy}) of the hydrodynamics. Comparing
(\ref{rate-ads}) and (\ref{rate-hydro}), we obtain
\begin{eqnarray}
S_{\infty}=\frac{4}{3}\frac{\rho_{0}}{T_{0}}
=
\left. \frac{4}{3}\frac{\rho \tau}{T}\right|_{\tau=\infty}.
\end{eqnarray}
This is nothing but the relationship among the entropy, the
energy (per unit rapidity and unit transverse area) and the
temperature obtained by thermodynamics at
$\tau=\infty$ where the system reaches thermal equilibrium.

Before closing this section,
let us give a technical remark to clarify the
meaning of the late time limit (\ref{Janik-limit}).
The readers might want to skip this paragraph at first reading.
One finds that the higher order terms we have neglected in
(\ref{expansion1}) and (\ref{expansion2}) contain
the terms proportional to
$v^{6}\tau^{-4/3}$ for example. On the other hand,
the leading order terms in $g_{\tau\tau}$
contains arbitrary higher power of $v$.
Therefore, neglect of the $O(\tau^{-4/3})$ terms is justified
only when $v$ is larger than $O(1)$.
In fact, we can justify the limit (\ref{Janik-limit})
when we extract the thermodynamic quantities of the system.
Such quantities are associated with
the horizon of the black hole and
the value of $v$ at the horizon is indeed $O(1)$ constant
at the late time. Now, we understand why the naive $\tau \to \infty$ limit
on the $z$-coordinate (the limit with fixing $z$ to be constant)
is not good for our purpose.
The position of the horizon grows with time:
$z_{0}\sim  \tau^{1/3}. $
If we treat $z$ to be a constant in the $\tau \to \infty$ limit,
the region we describe becomes infinitely far from the horizon.
In other words,
neglect of the terms of $(z/\tau^{1/3})^{n}$ ($n>0$) is not
justified around the horizon.
This means, a suitable coordinate is the $v$-coordinate
rather than the $z$-coordinate
to describe near the horizon at $\tau \to \infty$.

\section{Conditions on energy density and bulk singularity}
\label{singularity}

In Ref. \cite{Janik}, singularity analysis was used to select
a physical metric. Namely, starting with energy density
(without viscosity)\footnote{
The value of $l$ is restricted to be $0<l<4$ by
the positive energy condition for (\ref{T-diag-2}) \cite{Janik}.
}
\begin{eqnarray}
\rho=\frac{\rho_{0}}{\tau^{l}},
\end{eqnarray}
it is found that the late time bulk geometry is singular
except for a special value of $l$.
More precisely, $(R_{MNKL})^{2}$ at the order of $(\tau)^{0}$
has singularity at the horizon except for $l=\frac{4}{3}$,
the value for the perfect fluid.
In fluid dynamics, this value of $l$
is determined by the conservation
law and the equation of state.
However the bulk metric knows the correct form of the energy
density independently \cite{Janik}.
Therefore it is interesting to see whether requiring
the regularity of the metric at the horizon gives further control
over the behavior of the viscous term as well.
The late time bulk geometry with generic value of $\beta$
can be obtained through similar calculations starting with
the energy-momentum tensor (\ref{T-diag-2}) with $\rho$ given in
(\ref{solution}) or (\ref{critical-solution}).

If $\beta<1/3$, the viscous correction in $\rho$ is dominant
at the late time and $\rho\sim 1/\tau^{1+\beta}$ from (\ref{solution}).
This leads to the singular geometry since the proper time dependence
of the dominant term in $\rho$,
$1/\tau^{1+\beta}$, is not $1/\tau^{4/3}$.

If $\beta>1/3$,
the viscous correction is sub-leading and we should consider
$(R_{MNKL})^{2}$ to the sub-leading order.
The late time geometry is the same form
of (\ref{our-geometry}) except the following replacement
\begin{eqnarray}
\gamma\to \gamma'\equiv \eta_{0}/(\rho_{0} \tau^{\beta-1/3}),
\end{eqnarray}
and with $\rho$ given by (\ref{solution}).
We find that $(R_{MNKL})^{2}$ has the following structure:
\begin{eqnarray}
(R_{MNKL})^{2}
=\frac{8(5w^{16}+20w^{12}+174w^{8}+20w^{4}+5)}{(1+w^{4})^{4}}
+O(\tau^{-4/3}),
\label{R2}
\end{eqnarray}
where
\begin{eqnarray}
w\equiv
z \left( \frac{\rho}{3}\right)^{1/4}= \frac{z}{\tau^{1/3}}
 \left( \frac{\rho_{0}}{3}\right)^{1/4}\left(1+
    \frac{4 \gamma'}{1-3\beta} \right)^{1/4}.
\end{eqnarray}
In the absence of viscosity, the above result is reduced to
that of Ref. \cite{Janik}.
The first term in the right-hand side of (\ref{R2}), that contains
the viscous sub-leading corrections, is finite.
So the consideration of singularity does not give any further restriction to the viscosity term.
If $\beta>5/3$ the corrections due to the
viscosity-dependence give smaller effects in the metric than
the non-viscous $O(\tau^{-4/3})$ corrections which are already
discarded in the late time geometry.
So up to our approximation, the viscous effect is not visible
in this case.

For $\beta=1/3$ case, the viscous corrections are leading
order. We find that the metric (\ref{Janik-bulk}), where
$\frac{\rho_{0}}{3}\frac{1}{\tau^{4/3}}$ is
replaced with the second term of $\rho$ in
(\ref{critical-solution}), gives the
late time geometry\footnote{
In this case, the late time limit
should be taken by fixing $v=z (\log \tau)^{1/4}/\tau^{1/3}$.
}
as mentioned in Section \ref{non-vis}.
$(R_{MNKL})^{2}$ at the leading order is given in the same form
as that of (\ref{R2}) where $w$ is $w=z(\rho/3)^{1/4}$
with $\rho$ given by the second term of (\ref{critical-solution}).
There is no divergence at the leading order, although the value
of $\beta$ is not consistent with the hydrodynamics.

Altogether, our geometry is regular and the viscosity effect
is meaningful in the region of
\begin{eqnarray}
1/3 \le  \beta< 5/3.
\end{eqnarray}
Indeed, $\beta=1$ is within  this region.
This means that in the late time geometry (or the late time
fireball dynamics),
the viscous term gives visible contribution to the dynamics of
the fireball.

\section{Conclusions}

We considered the gravity dual of large-$N_{c}$ ${\cal N}=4$ SYM
fluid undergoing one dimensional expansion with account of
shear viscosity.
We obtained the late time bulk geometry to the minimal
order of the viscous corrections in the analytic form.
We found that our viscous corrections do not break
the regularity within our approximation.
We also found that the time evolution of the thermodynamic
quantities given by the late time geometry is
consistent with the hydrodynamic analyses.

We saw that
the holographic dual of the hydrodynamics contains much more
information than the hydrodynamics, since AdS/CFT already
contains essential information of microscopic gauge theory dynamics.
For example, the holographic dual of the hydrodynamics gave a
simple derivation of Stefan-Boltzmann's law in the strongly
coupled region with precise Stefan-Boltzmann constant.
The integration constant in the hydrodynamic
equation was also given by looking at the horizon of the dual
geometry.

We believe that by probing the resulting geometry, one can
extract many of information of strongly interacting
system in principle.
It is important to compute various physical quantities based
on the obtained geometry \cite{inprogress}. We can also consider
various extensions of the present work. Inclusion of the
higher-order corrections, generalization to the systems with chemical
potential, consideration of the systems with three-dimensional
expansion are possible directions.
One can also consider the effects of the bulk viscosity whose
presence violates the conformal invariance.
In the real QCD, conformal invariance must be broken and
including it might be relevant for more realistic account of
RHIC fireball.
We hope that the present work will shed light on AdS/CFT for
non-static non-equilibrium systems and  holographic description of
RHIC physics.

\vspace{0.5cm}

\noindent
{\large\bf Acknowledgments}\\
We would like to thank I. Zahed,  M. Natsuume, K. Furuuchi, Youngman
Kim, Ho-Ung Yee and S. Hirano for discussions, and Sang Pyo Kim for
reading the manuscript carefully. We would also like to thank the
APCTP where a part of this work was done. This work was supported by
the SRC Program of  the KOSEF through the Center for Quantum
Space-time (CQUeST) of Sogang University with grant number R11 - 2005
- 021 and also by KOSEF Grant R01-2004-000-10520-0.

\appendix
\section{Derivation of the metric (\ref{our-geometry})}
\label{derivation}

First, we find the following expressions by solving the Einstein's
equation recursively  to the order of $v^{28}$:
\begin{eqnarray}
g_{\tau\tau}(\tau,v)
&=&-1+3a-4a^{2}+4a^{3}-4a^{4}+4a^{5}-4a^{6}+4a^{7}+\cdots
\nonumber \\
&&
-\frac{2\eta_{0}v^{4}}{\tau^{2/3}}
\left(
1+\frac{4a}{3}
\left(
-2+3a-4a^{2}+5a^{3}-6a^{4}+7a^{5}+\cdots
\right)
\right)
+O(\tau^{-4/3}),
\nonumber \\
\frac{g_{yy}(\tau,v)}{\tau^{2}}
&=&1+a
-\frac{2\eta_{0}v^{4}}{\tau^{2/3}}
\left(
1+\frac{2a}{3}
\left(
1+\frac{a}{3}+\frac{a^{2}}{3}+\frac{a^{3}}{5}+\frac{a^{4}}{5}
+\frac{a^{5}}{7}+\cdots
\right)
\right)
+O(\tau^{-4/3}),
\nonumber \\
g_{xx}(\tau,v)
&=&1+a
+
\frac{2\eta_{0}v^{4}}{\tau^{2/3}}
\frac{a}{3}
\left(
1+\frac{a}{3}+\frac{a^{2}}{3}+\frac{a^{3}}{5}+\frac{a^{4}}{5}
+\frac{a^{5}}{7}+\cdots
\right)
+O(\tau^{-4/3}),
\end{eqnarray}
where
$a= {\rho_{0} v^{4}}/{3}.
$

The power series in the right-hand sides are re-summed to give
the analytic form of the metric:
\begin{eqnarray}
g_{\tau\tau}(\tau,v)
&=&
-\frac{(1-\frac{\rho_{0} v^{4}}{3})^{2}}{1+\frac{\rho_{0} v^{4}}{3}}
-\frac{2\eta_{0}v^{4}}{3\tau^{2/3}}
\frac{(1-\frac{\rho_{0} v^{4}}{3})(3+\frac{\rho_{0} v^{4}}{3})}
{(1+\frac{\rho_{0} v^{4}}{3})^{2}}+O(\tau^{-4/3}),
\nonumber \\
\frac{g_{yy}(\tau,v)}{\tau^{2}}
&=&
1+\frac{\rho_{0} v^{4}}{3}
-\frac{2\eta_{0}v^{4}}{3\tau^{2/3}}
\left(
1+
\frac{1+\frac{\rho_{0} v^{4}}{3}}{\frac{\rho_{0} v^{4}}{3}}
\log
\left(
\frac{1+\frac{\rho_{0} v^{4}}{3}}{1-\frac{\rho_{0} v^{4}}{3}}
\right)
\right)
+O(\tau^{-4/3}),
\nonumber \\
g_{xx}(\tau,v)
&=&
1+\frac{\rho_{0} v^{4}}{3}
-\frac{2\eta_{0}v^{4}}{3\tau^{2/3}}
\left(
1-\frac{1}{2}
\frac{1+\frac{\rho_{0} v^{4}}{3}}{\frac{\rho_{0} v^{4}}{3}}
\log
\left(
\frac{1+\frac{\rho_{0} v^{4}}{3}}{1-\frac{\rho_{0} v^{4}}{3}}
\right)
\right)
+O(\tau^{-4/3}),
\label{g-correct}
\end{eqnarray}
We can check explicitly that (\ref{g-correct}) is indeed
the solution of the Einstein's equation that is accurate
to the order of $\tau^{-2/3}$.
The important feature of (\ref{g-correct}) is that all the
terms at the order of $\tau^{-2/3}$ are proportional to $\eta_{0}$.
(\ref{g-correct}) is rewritten in the following forms
up to  $O(\gamma^{2})$ terms:
\begin{eqnarray}
g_{\tau\tau}(\tau,v)
&=&
-\frac{(1-\frac{\rho_{0}(1-2\gamma) v^{4}}{3})^{2}}
{1+\frac{\rho_{0}(1-2\gamma) v^{4}}{3}}
+O(\tau^{-4/3}),
\nonumber \\
\frac{g_{yy}(\tau,v)}{\tau^{2}}
&=&
\left(1+\frac{\rho_{0}(1-2\gamma) v^{4}}{3}\right)
\left(
1-2\gamma
\log
\left(
\frac{1+\frac{\rho_{0}(1-2\gamma) v^{4}}{3}}
{1-\frac{\rho_{0}(1-2\gamma) v^{4}}{3}}
\right)
\right)
+O(\tau^{-4/3}),
\nonumber \\
g_{xx}(\tau,v)
&=&
\left(1+\frac{\rho_{0}(1-2\gamma) v^{4}}{3}\right)
\left(
1+\gamma
\log
\left(
\frac{1+\frac{\rho_{0}(1-2\gamma) v^{4}}{3}}
{1-\frac{\rho_{0}(1-2\gamma) v^{4}}{3}}
\right)
\right)
+O(\tau^{-4/3}),
\label{g-guess-ap}
\end{eqnarray}
or more simply,
\begin{eqnarray}
g_{\tau\tau}(\tau,v)
&=&
-\frac{(1-\frac{\rho_{0}(1-2\gamma) v^{4}}{3})^{2}}
{1+\frac{\rho_{0}(1-2\gamma) v^{4}}{3}}
+O(\tau^{-4/3}),
\nonumber \\
\frac{g_{yy}(\tau,v)}{\tau^{2}}
&=&
\left(1+\frac{\rho_{0}(1-2\gamma) v^{4}}{3}\right)
\left(
\frac{1+\frac{\rho_{0}(1-2\gamma) v^{4}}{3}}
{1-\frac{\rho_{0}(1-2\gamma) v^{4}}{3}}
\right)^{-2\gamma}
+O(\tau^{-4/3}),
\nonumber \\
g_{xx}(\tau,v)
&=&
\left(1+\frac{\rho_{0}(1-2\gamma) v^{4}}{3}\right)
\left(
\frac{1+\frac{\rho_{0}(1-2\gamma) v^{4}}{3}}
{1-\frac{\rho_{0}(1-2\gamma) v^{4}}{3}}
\right)^{\gamma}
+O(\tau^{-4/3}).
\label{g-guess-ap2}
\end{eqnarray}
The differences among (\ref{g-correct}), (\ref{g-guess-ap}) and
(\ref{g-guess-ap2}) are at the order of $O(\tau^{-4/3})$
(that is the same order of $O(\gamma^{2})$) and
they share the same terms to the order of $\gamma$.

\section{Second order formalism of dissipative
relativistic hydrodynamics}
\label{second-from}

In this appendix, we briefly introduce the causal dissipative
relativistic hydrodynamics that is also referred to the
second order dissipative relativistic hydrodynamics.
We shall show that the first order formalism we have employed
gives a good approximation of the second order formalism at the
late time in our setup.

It is known that the first order formalism
of the dissipative relativistic hydrodynamics has a problem;
the viscous and the thermal signal propagates
instantaneously and causality is broken.
(See for example, \cite{Muronga}.)
In the second order formalism \cite{IS}, the relaxation time of
the fluid is introduced to maintain the causality.
The energy density evolution equation in our setting is
\cite{Muronga}:
\begin{eqnarray}
\frac{d\rho}{d\tau}
=-\frac{4}{3}\frac{\rho}{\tau}+\frac{\Phi}{\tau},
\end{eqnarray}
where
\begin{eqnarray}
\Phi &=& \frac{4}{3}\frac{\eta}{\tau}
\:\:\:\:\:\:\:\:
\:\:\:\:\:\:\:\:(\mbox{\rm the first order formalism}),\\
\tau_{\pi}\frac{d\Phi}{d\tau}
&=&
-\Phi+\frac{4}{3}\frac{\eta}{\tau}
\:\:\:\:\:(\mbox{\rm the second order formalism}),
\label{second}
\end{eqnarray}
and $\tau_{\pi}$ is the relaxation time of the system.
Note that the $\tau_{\pi} \to 0$ limit
gives the first order formalism.

Let us evaluate the difference between the first
order formalism and the second order formalism for our case.
We begin with the assumption that the proper time dependence of
the shear viscosity is given by (\ref{eta}).
Substituting (\ref{eta}) to (\ref{second}), we find
\begin{eqnarray}
\Phi&=&
\frac{4}{3}\frac{\eta}{\tau}
\left\{
1+(1+\beta)\frac{\tau_{\pi}}{\tau}
+(1+\beta)(2+\beta)\left(\frac{\tau_{\pi}}{\tau}\right)^{2}
+\cdots
\right\}
\nonumber \\
&+& \mbox{constant} \times e^{-\tau/\tau_{\pi}}.
\end{eqnarray}
This means that the second order formalism approaches to
the first order formalism when
\begin{eqnarray}
\frac{\tau_{\pi}}{\tau} \ll 1.
\label{first-order-cond}
\end{eqnarray}
Therefore, our analyses
based on the first order formalism (\ref{diff-eq}) with
assumption (\ref{eta}) and the late time
approximation are self-consistent.
The condition (\ref{first-order-cond}) also agrees with our
basic assumption that the microscopic time scale $\tau_{\pi}$
is short enough comparing to the macroscopic time scale so that
the hydrodynamic description is valid.
Small value of $\tau_{\pi}$ also matches the fact that our fluid
consists of strongly interacting particles.

\section{Consistency of $\eta=\eta_{0}/\tau$ with $\eta\sim T^{3}$}
\label{consistency}

Let us check self-consistency of our assumption (\ref{beta1}).
Starting with $\eta=\eta_{0}/\tau$,
the energy density is given as
\begin{eqnarray}
\rho(\tau)
=
\frac{\rho_{0}}{\tau^{4/3}}
\left(
1-2\gamma
\right),
\label{e-density}
\end{eqnarray}
where
\begin{eqnarray}
\gamma \equiv \frac{\eta_{0}}{\rho_{0}\tau^{2/3}}.
\label{def-gamma}
\end{eqnarray}
The relationship $\eta \propto T^{3} \propto \rho^{3/4}$
makes further corrections to the shear viscosity like
\begin{eqnarray}
\eta=
\frac{\eta_{0}}{\tau}
\left\{1+O(\gamma) \right\}
,
\label{eta-expand}
\end{eqnarray}
and the $O(\gamma \tau^{-1})$ correction in (\ref{eta-expand})
makes further corrections to the energy density recursively.
However, all such corrections are at the higher order of $\gamma$
and we can neglect them if $\gamma \ll 1$.
Let us define our approximation precisely:
\begin{itemize}
  \item We consider only the region of $\gamma \ll 1$.
  \item We consider $\eta \tau$ to the order of $1$
and $\rho \tau^{4/3}$ to the order of $\gamma$.
In other words, we consider only to the order of $\eta_{0}$.
\end{itemize}
The above makes our framework to be self-consistent.\footnote{
The physical meaning of $\gamma \ll 1$ is that the contribution
of the shear viscosity to the energy density is small.
This is also a necessary condition to justify the hydrodynamic
treatment of the system.}
Note that the positive energy condition
for (\ref{e-density}) is also guaranteed by the above approximation.


\begin{thebibliography}{99}

\newcommand{\J}[4]{{\sl #1} {\bf #2} (#3) #4}
\newcommand{\andJ}[3]{{\bf #1} (#2) #3}
\newcommand{\AP}{Ann.\ Phys.\ (N.Y.)}
\newcommand{\MPL}{Mod.\ Phys.\ Lett.}
\newcommand{\NP}{Nucl.\ Phys.}
\newcommand{\PL}{Phys.\ Lett.}
\newcommand{\PR}{Phys.\ Rev.}
\newcommand{\PRL}{Phys.\ Rev.\ Lett.}
\newcommand{\ATMP}{Adv.\ Theor.\ Math.\ Phys.}
\newcommand{\JHEP}{JHEP}
\newcommand{\IJMP}{Int.\ J.\ Mod.\ Phys.}
\newcommand{\JETPL}{JETP\ Lett.}
\newcommand{\SJNP}{Sov.\ J.\ Nuc.\ Phys.}
\newcommand{\PTP}{Prog.\ Theor.\ Phys.}

\bibitem{AdS/CFT}
J. M. Maldacena,
``The Large N Limit of Superconformal Field Theories and
Supergravity'',
{\sl Adv. Theor. Math. Phys.} {\bf 2} (1998) 231,
{\sl Int. J. Theor. Phys.} {\bf 38} (1999) 1113,
hep-th/9711200;\\
S. S. Gubser, I. R. Klebanov and A. M. Polyakov,
``Gauge Theory Correlators from Non-Critical String Theory'',
\J{\PL}{B428}{1998}{105},
hep-th/9802109;\\
E. Witten,
``Anti De Sitter Space and Holography'',
{\sl Adv. Theor. Math. Phys.} {\bf 2} (1998) 253,
hep-th/9802150.

\bibitem{RHIC}
E. V. Shuryak,
``What RHIC Experiments and Theory tell us about Properties
of Quark-Gluon Plasma?'',
\J{\NP}{A750}{2005}{64},
hep-ph/0405066;
M. J. Tannenbaum,
``Recent results in relativistic heavy ion collisions:
from ``a new state of matter'' to "the perfect fluid"'',
{\sl Rept. Prog. Phys.} {\bf 69} (2006) 2005,
nucl-ex/0603003.

\bibitem{SZ}
S.~J.~Sin and I.~Zahed,
``Holography of radiation and jet quenching,''
{\sl Phys.\ Lett.}\ {\bf B608} (2005) 265,
hep-th/0407215.

\bibitem{Nastase}
H. Nastase,
``The RHIC fireball as a dual black hole'',
hep-th/0501068;
H. Nastase,
``DBI skyrmion, high energy (large s) scattering and
fireball production'',
hep-th/0512171;
H. Nastase,
``More on the RHIC fireball and dual black holes'',
hep-th/0603176.

\bibitem{SSZ}
E. Shuryak, S-J. Sin and I. Zahed,
``A Gravity Dual of RHIC Collisions'',
hep-th/0511199.

\bibitem{Aharony}
O. Aharony, S. Minwalla and Toby Wiseman,
``Plasma-balls in large N gauge theories and localized black holes'',
{\sl Class. Quant. Grav.} {\bf 23} (2006) 2171,
hep-th/0507219.

\bibitem{AdS/QCD}
For earlier attempts for
QCD-like models, see for example the following reviews and the
references therein:\\
O. Aharony,
``The non-AdS/non-CFT correspondence, or three different paths to QCD'',
hep-th/0212193;
A. Zaffaroni,
``RTN lectures on the non AdS/non CFT Correspondence'',
PoS(RTN2005)005 at {\tt http://pos.sissa.it/}.\\
More recent proposals are:\\
T. Sakai and S. Sugimoto,
``Low energy hadron physics in holographic QCD'',
\J{\PTP}{113}{2005}{843},
hep-th/0412141;
T. Sakai and S. Sugimoto,
``More on a holographic dual of QCD'',
\J{\PTP}{114}{2006}{1083},
hep-th/0507073;
J.~Erlich, E.~Katz, D.~T.~Son and M.~A.~Stephanov,
``QCD and a holographic model of hadrons,''
{\sl Phys.\ Rev.\ Lett.}\  {\bf 95} (2005) 261602,
hep-ph/0501128;
L.~Da Rold and A.~Pomarol,
``Chiral symmetry breaking from five dimensional spaces,''
{\sl Nucl.\ Phys.}\ {\bf B721} (2005) 79,
hep-ph/0501218.


\bibitem{Janik}
R. A. Janik and R. Peschanski,
``Asymptotic perfect fluid dynamics as a consequence of AdS/CFT'',
\J{\PR}{D73}{2006}{045013},
hep-th/0512162.

\bibitem{Janik-2}
R. A. Janik and R. Peschanski,
``Gauge/gravity duality and thermalization of
a boost-invariant perfect fluid'',
\J{\PR}{D74}{2006}{046007},
hep-th/0606149.
\bibitem{HSS}
S. de Haro, K. Skenderis and S. N. Solodukhin,
``Holographic Reconstruction of Spacetime and Renormalization
in the AdS/CFT Correspondence'',
{\sl Commun. Math. Phys.} {\bf 217} (2001) 595,
hep-th/0002230.

\bibitem{Skend}
K. Skenderis,
``Lecture Notes on Holographic Renormalization'',
{\sl Class. Quant. Grav.} {\bf 19} (2002) 5849,
hep-th/0209067.

\bibitem{Bjorken}
J. D. Bjorken,
``Highly relativistic nucleus-nucleus collisions:
The central rapidity region",
\J{\PR}{D27}{1983}{140}.

\bibitem{KH-Chau}
P. F. Kolb and U. Heinz,
``Hydrodynamic description of ultrarelativistic heavy-ion collisions'',
nucl-th/0305084.


\bibitem{bulk-zero}
E. Shuryak,
``Why does the Quark-Gluon Plasma at RHIC behave as a nearly
ideal fluid?'',
{\sl Prog. Part. Nucl. Phys.} {\bf 53} (2004) 273,
hep-ph/0312227.

\bibitem{PSS-1}
G. Policastro, D. T. Son and A. O. Starinets,
``Shear viscosity of strongly coupled N=4 supersymmetric
Yang-Mills plasma'',
\J{\PRL}{87}{2001}{081601},
hep-th/0104066.



\bibitem{Muronga}
A. Muronga,
``Causal Theories of Dissipative Relativistic Fluid Dynamics
for Nuclear Collisions'',
\J{\PR}{C69}{2004}{034903},
nucl-th/0309055.




\bibitem{GKPeet}
S. S. Gubser, I. R. Klebanov and A. W. Peet,
``Entropy and Temperature of Black 3-Branes'',
\J{\PR}{D54}{1996}{3915},
hep-th/9602135.

\bibitem{IS}
W. Israel,
``Nonstationary irreversible thermodynamics:
A Causal relativistic theory'',
{\sl Ann. Phys.} {\bf 100} (1976) 310;
W. Israel and J. M. Stewart,
``Transient relativistic thermodynamics and kinetic theory'',
{\sl Ann. Phys.} {\bf 118} (1979) 341.

\bibitem{inprogress}
S. Nakamura and S-J. Sin,
work in progress.


\end{thebibliography}
\end{document}